\documentclass{aastex63}
\usepackage{natbib}
\usepackage{amsmath}

\begin{document}
	
	\title{Transit Light-curves for Exomoons: Analytical Formalism}
	
	\author[0000-0001-8018-0264]{Suman Saha}
	\affiliation{Indian Institute of Astrophysics, II Block, Koramangala, Bengaluru, India}
	\affiliation{Pondicherry University, R.V. Nagar, Kalapet, Puducherry, India}
	
	\correspondingauthor{Suman Saha}
	\email{suman.saha@iiap.res.in}
	
	\author[0000-0002-6176-3816]{Sujan Sengupta}
	\affiliation{Indian Institute of Astrophysics, II Block, Koramangala, Bengaluru, India}
	
	\accepted{for publication in The Astrophysical Journal}
	
	\begin{abstract}
		
	The photometric transit method has been the most effective method to detect and characterize exoplanets as several ground-based as well as space-based survey missions have discovered thousands of exoplanets using this method. With the advent of the upcoming next generation large telescopes, the detection of exomoons in a few of these exoplanetary systems is very plausible. In this paper, we present a comprehensive analytical formalism in order to model the transit light curves for such moon hosting exoplanets. In order to achieve analytical formalism, we have considered circular orbit of the exomoon around the host planet, which is indeed the case for tidally locked moons. The formalism uses the radius and orbital properties of both the host planet and its moon as model parameters. The co-alignment or non-coalignment of the orbits of the planet and the moon is parameterized using two angular parameters and thus can be used to model all the possible orbital alignments for a star-planet-moon system. This formalism also provides unique and direct solutions to every possible star-planet-moon three circular body alignments. Using the formula derived, a few representative light curves are also presented.
		
	\end{abstract}
	
	
	\section{Introduction}\label{sec1}
		
	The large number of natural satellites around the planets in our Solar system suggests a high possibility of the existence of such sub-planetary bodies around many of the exoplanets discovered till date.  In the past two decades, more than four thousand exoplanets with a wide range of size and mass have been discovered by using various detection techniques. However, the discovery of natural satellites (also known as exomoons) in those systems still remains elusive.
	
	Out of the various detection techniques used for the discovery of exoplanets, the transit method has been proven to be the most effective. Apart from detection, the transit method also provides a way to estimate the size and orbital properties of the exoplanets accurately. Derivatives of the transit method involving the effect of the exomoon on the companion exoplanet, such as the Transit-Timing Variation \citep[TTV,][]{1999A&AS..134..553S, 2006A&A...450..395S}, and the Transit-Duration Variation \citep[TDV,][]{2009MNRAS.392..181K} have been proposed for the detection of exomoons. However, the amplitude of these effects is extremely small for sub-Earth mass exomoons and so far no confirmed exomoon candidate has been detected using these techniques \citep{fox2021exomoon, 2020ApJ...900L..44K, 2021MNRAS.500.1851K}. Several other techniques have also been proposed for the detection of exomoons, such as photometric orbital sampling effect \citep{2014ApJ...787...14H, 2018AJ....155...36T}, imaging of mutual transits \citep{2007A&A...464.1133C}, microlensing \citep{2002ApJ...580..490H}, spectroscopy \citep{2004AsBio...4..400W, 2006PASP..118.1136J, 2019ApJ...885..168O}, polarimetry of self-luminous exoplanets \citep{2016ApJ...824...76S}, doppler monitoring of directly images exoplanets \citep{2015ApJ...812....5A}, pulsar timing \citep{2008ApJ...685L.153L} and radio emissions of giant exoplanets \citep{2014ApJ...791...25N, 2016ApJ...821...97N}. However, no confirmed exomoon candidate has yet been detected by using any of these techniques.
	
	A shortcoming of the transit method is that the transit probability decreases with the increase in the orbital distance of the exoplanets from their host-stars. Also, with the increase in the orbital distance, the orbital period of the exoplanets increases reducing the probability of detection as it requires continuous monitoring for a longer time-period to confirm the detection. Both these factors have severely constrained the discovery of exoplanets in wider orbits. Previous studies \citep{2010ApJ...719L.145N, 2016ApJ...817...18S, 2021PASP..133i4401D} has shown that the exoplanets in close-in orbits are likely to lose any natural satellite during the orbital migration. This could be the prime reason behind the lack of discovery of exomoons around the planetary systems discovered till date. With the installation of  dedicated survey telescopes, both ground-based and space-based, and a combination of the observations from multiple observing facilities can enable the detection of exoplanets in wider orbits in near future. Such facilities may also enable to detect the natural satellites around the exoplanets.
	
	Another major factor that makes it difficult to detect the exomoons through photometric transit method is the requirement of extreme precision. Survey missions like the CoRoT, Kepler and TESS, and the 2m class Hubble Space Telescope have made it possible to detect and study exoplanets as small as the Earth. However, the detection probability of such smaller exoplanets is more around the smaller late-K or M dwarf stars compared to the larger stars of similar apparent magnitude.
	This is because the precision required to detect exoplanets is proportional to the ratio of disc area of the planet to the star. On the other hand, the largest of the natural satellites in our solar system has a radius even smaller than half of the radius of the Earth. Such smaller natural satellites may be abundant in many exoplanetary systems, but their detection would require a much better photometric precision. All these factors sum up to the inference that the detection of exomoons would require a precision higher than that achievable using the currently available instruments. However, the next generation large telescopes, such as the James Webb Space Telescope (JWST), the European Extremely Large Telescope (E-ELT), the Thirty Meter Telescope (TMT), and the Giant Magellan Telescope (GMT) etc. will make it possible to achieve such extremely high precision. Also, the major improvements in small-scale noise reduction techniques \citep{2015ApJ...810L..23J, 2019AJ....157..102L, 2019AJ....158...39C, 2021AJ....162...18S, 2021AJ....162..221S} will help in such studies to improve the precision of photometric lightcurves.
	
	As a consequence, it's very much likely that the improved facilities in near future will make it possible to detect the exomoons through photometric transit method. In that case, a self-consistent and mathematically straightforward analytical formalism will be necessary to model the transit lightcurves of stars having moon hosting planets and to estimate the physical properties of the exomoons. The theoretical models need also to be unambiguous enough to be applicable for all possible realistic scenarios of the position of the exomoon with respect to the host exoplanet, and will thus help to strategize the observational parameters to such extremely time-critical observations. We therefore, present in this paper a comprehensive and generic analytical formalism for the lightcurves of a transiting exoplanetary system with a exomoon in terms of the radius and orbital parameters of both the planet and the moon. An important aspect of such formalism is to formulate the motion of the three bodies in a common reference frame with respect to the observer. The existing formalism \citep{2011MNRAS.416..689K, 2018SciA....4.1784T} rely on the rotational transformation of the individual orbits for this purpose. On the other hand, we present a comparatively simple and straightforward analytical formalism by taking into account only the physically significant orbital parameters of the exoplanet and a circularly orbiting exomoon, such that one can easily model the transit light curves for every possible orbital alignments of the system. To account for the independent spacial inclination of the orbits of the planet and the moon in a simpler way, we have used a two angular parameter approach. We have also provided direct and straightforward solutions to the conditions for various alignments of the star-planet-moon system, especially for the cases where all the three circular bodies intersect with each other. In section~\ref{sec2}, we discuss the analytical formalism; in section~\ref{sec3}, we present the results; and in section~\ref{sec4}, we conclude our study.
	
	\section{Analytical formalism}\label{sec2}
	
	Let's consider a natural satellite or moon with radius $r_m$, orbiting a planet with radius $r_p$, where $r_m$ and $r_p$ are expressed in term of the radius of the star. If the moon is sufficiently massive, the barycenter of the planet-moon system will be significantly away from the center of the planet even though it lies inside the planetary surface. However, the barycenter will follow the same orbit as the planet would in the absence of a moon. In order to achieve a simple analytical formalism, we have considered a circular orbit for the barycenter around the star. This provides us the advantage of the underlying symmetry. For most of the cases, the orbital eccentricity of an exoplanet can only be estimated by using the radial-velocity method, and a prior knowledge of it would be required to model the transit lightcurve correctly. However, in the absence of a prior information of the eccentricity, by modeling the transit signal for a circular orbit would result in a slightly different value for the orbital semi-major axis and the orbital inclination of the planet. However, it wouldn't affect the orbital properties of the moon. 
	
	In the formalism presented here, we have used subscripts s, p, m and b to denote the star, the planet, the moon and the planet-moon barycenter respectively. Now, the separation of the barycenter of the planet-moon system from the center of the star is given by,
	
	\begin{equation}
		z_{sb} = a_b \sqrt{\sin^2\theta_b + \cos^2\theta_b \cos^2 i_b}
	\end{equation}
	
	\begin{equation}
		\theta_b = \frac{2 \pi}{P_b} (t-t_{0b})
	\end{equation}
	
	where $a_b$ is the semi-major axis; $i_b$ is the inclination angle; $\theta_b$ is the orbital phase; $P_b$ is the orbital period; and $t_{0b}$ is the mid-transit time of the planet-moon barycenter around the star.
	
	\begin{figure}
		\centering
		\includegraphics[width=\linewidth]{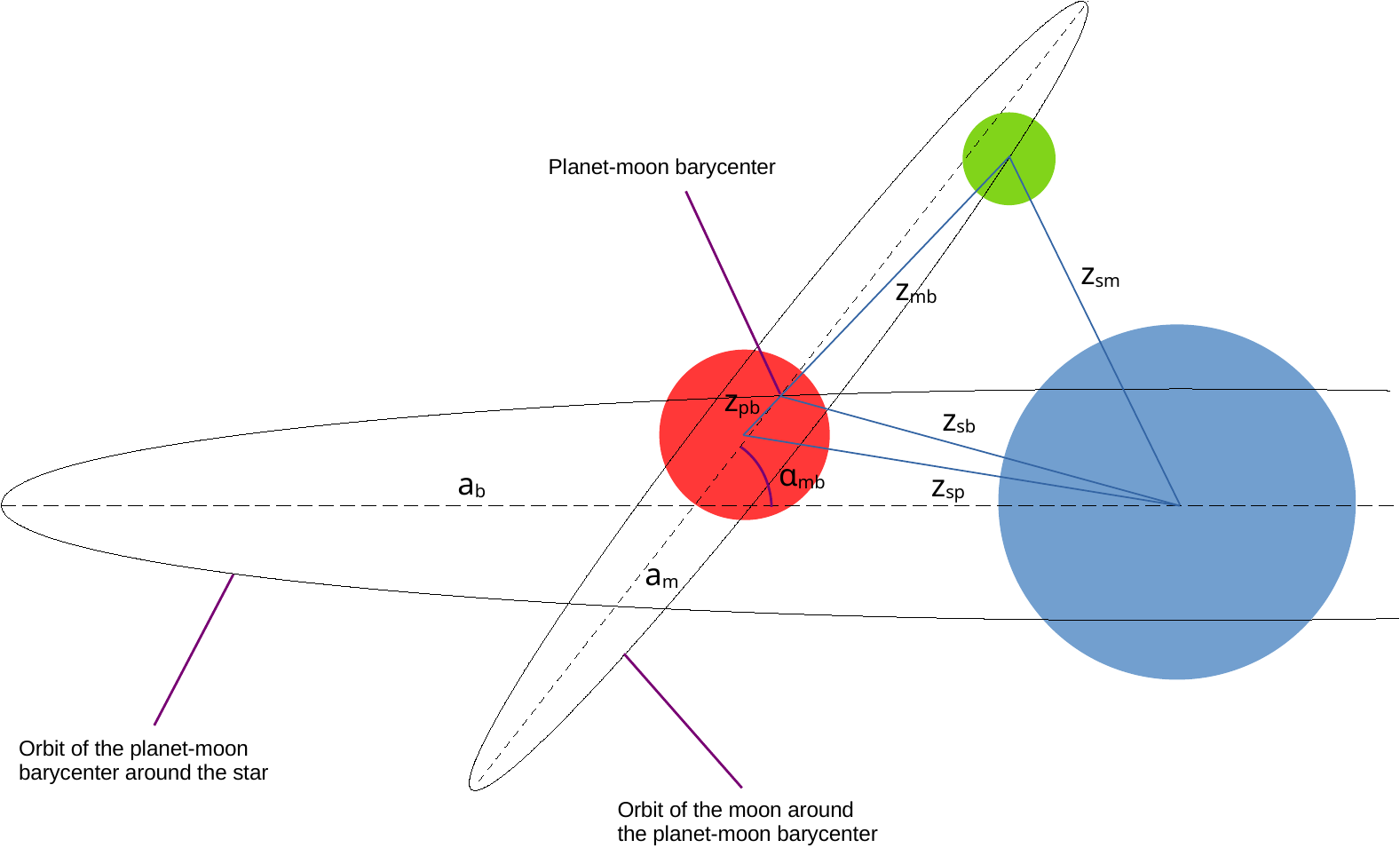}
		\caption{Orbital orientation of the star-planet-moon system from the observer's point of view, showing $z_{sp}$, $z_{pm}$ and $z_{sm}$, the separations between the centers of the star and the planet, the planet and the moon, and the star and the moon respectively; $z_{sb}$, $z_{pb}$ and $z_{mb}$, the separation of the planet-moon barycenter from the centers of the star, the planet and the moon respectively; $\alpha_{mb}$, the angle between the major axes of the projected orbits of the planet-moon barycenter around the star and the moon around the planet-moon barycenter}; $a_b$ and $a_m$, the orbital semi-major axes of the planet-moon barycenter around the star and the moon around the planet-moon barycenter respectively.
		\label{fig:fig1}
	\end{figure}
	
	If we consider the distance between centers of the moon and the planet to be $r_{pm}$, the distance of the center of the moon from the planet-moon barycenter is $a_m = r_{pm}/(1+M_m/M_p)$, where $M_m$ is the mass of the moon and $M_p $ is the mass of the planet; and the distance of the center of the planet from the planet-moon barycenter is $a_p = r_{pm} - a_m$. Now, the separation between the center of the moon and the barycenter of the planet-moon system is given by,
	
	\begin{equation}
		z_{mb} = a_m \sqrt{\sin^2\theta_m + \cos^2\theta_m \cos^2 i_m}
	\end{equation}
	
	\begin{equation}
		\theta_m = \frac{2 \pi}{P_m} (t-t_{0m})
	\end{equation}
	
	and, the separation of the center of the planet from the barycenter of the planet-moon system can be written as
	
	\begin{equation}
		z_{pb} = a_p \sqrt{\sin^2\theta_m + \cos^2\theta_m \cos^2 i_m}
	\end{equation}

	where $a_m$ is the semi-major axes of the moon, $a_p$ is the semi-major axes of the planet; $i_m$ is the inclination angle; $\theta_m$ is the orbital phase; $P_m$ is the orbital period; and $t_{0m}$ is the mid-transit time of the moon around the planet-moon barycenter. Note that, the orbital inclination angle of the moon is also measured with respect to the point of view of the observer. The separation between the center of the planet and the center of the moon is,
	
	\begin{equation}
		z_{pm} = z_{mb} + z_{pb}
	\end{equation}
	
	We denote the angle between the major axes of the projected orbits of the planet-moon barycenter around the star and the moon around the planet-moon barycenter as $\alpha_{mb}$  (see Figure \ref{fig:fig1}). Now, the separation between the centers of the planet and the star can be written as,
	
	\begin{equation}
		z_{sp} = \sqrt{z_{sb}^2 + z_{pb}^2 -2z_{sb}z_{pb}\cos \phi}
	\end{equation}
	
	\begin{equation}
		\phi = \alpha_{mb} + \eta - \eta_1
	\end{equation}
	
	\begin{equation}
		\eta = 
		\begin{cases}
			\cos^{-1} \frac{\cos \theta_b \cos i_b}{\sqrt{\sin^2\theta_b + \cos^2\theta_b \cos^2 i_b}}, & 0\le \theta_b \le \pi \\
			-\cos^{-1} \frac{\cos \theta_b \cos i_b}{\sqrt{\sin^2\theta_b + \cos^2\theta_b \cos^2 i_b}}, & -\pi\le \theta_b<0
		\end{cases}
	\end{equation}
	
	\begin{equation}
		\eta_1 = 
		\begin{cases}
			\cos^{-1} \frac{\cos \theta_m \cos i_m}{\sqrt{\sin^2\theta_m + \cos^2\theta_m \cos^2 i_m}}, & 0\le \theta_m \le \pi \\
			-\cos^{-1} \frac{\cos \theta_m \cos i_m}{\sqrt{\sin^2\theta_m + \cos^2\theta_m \cos^2 i_m}}, & -\pi\le \theta_m<0
		\end{cases}
	\end{equation}

	where $\phi$ is the angle between $z_{sb}$ and $z_{pb}$. Similarly, the separation between the centers of the moon and the star is written as,
	
	\begin{equation}
		z_{sm} = \sqrt{z_{sb}^2 + z_{mb}^2 -2z_{sb}z_{mb}\cos \phi_1}
	\end{equation}
	
	\begin{equation}
		\phi_1 = \pi - \phi
	\end{equation}
	
	If the ratio between the mass of the moon and the planet is assumed very small, the barycenter of the planet-moon system could be approximated at the center of the planet, i.e. $a_p = 0$, in which case the model simplifies to $z_{pb} = 0$, $z_{sp} = z_{sb}$ and $z_{pm} = z_{mb}$.
	
	\begin{figure}
		\centering
		\includegraphics[width=0.9\linewidth]{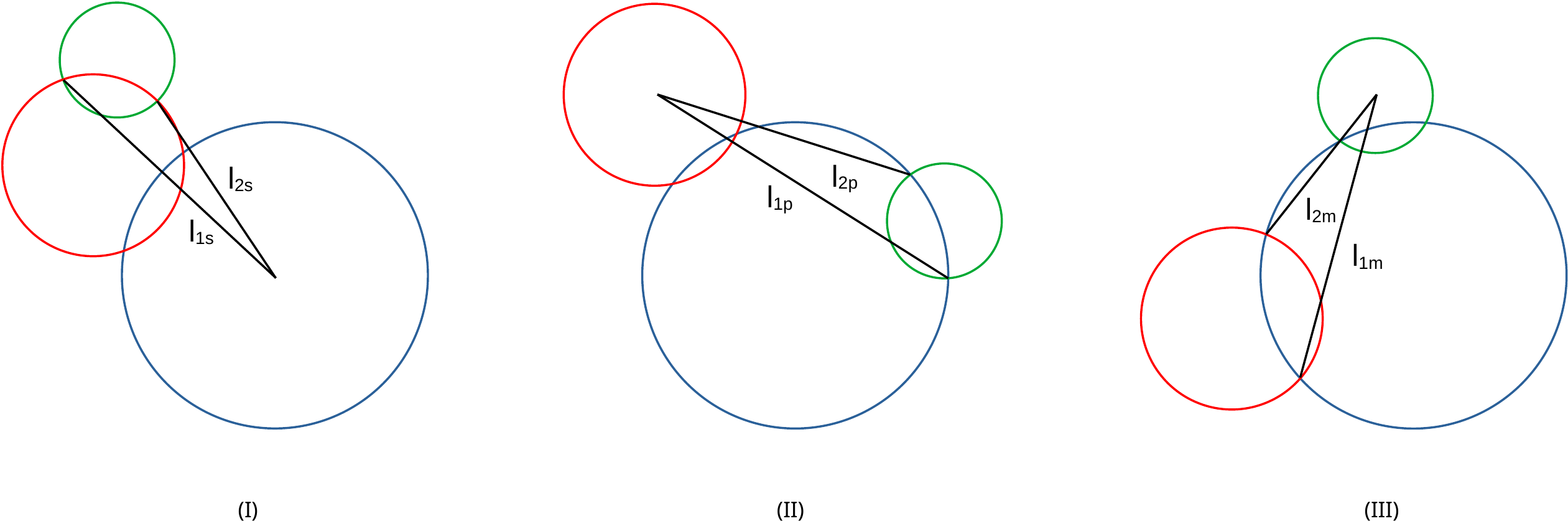}
		\caption{(I) Alignment with star-planet and planet-moon intersections showing $l_{1s}$ and $l_{2s}$, the separation of the star from the points of intersection of the planet and the moon; (II) Alignment with star-planet and star-moon intersections showing $l_{1p}$ and $l_{2p}$, the separation of the planet from the points of intersection of the star and the moon; (III) Alignment with star-planet and star-moon intersections showing $l_{1m}$ and $l_{2m}$, the separation of the moon from the points of intersection of the star and the planet.}
		\label{fig:fig11}
	\end{figure}
	
	The separation of the star from the points of intersection of the moon and the planet (see Figure \ref{fig:fig11} and Appendix \ref{sec7}) are given by,
	
	\begin{equation}
		l_{1s} = \sqrt{r_p^2+z_{sp}^2-2r_pz_{sp}\cos \bigg(\cos^{-1} \frac{z_{sp}^2+z_{pm}^2-z_{sm}^2}{2z_{sp}z_{pm}} + \cos^{-1} \frac{r_p^2-r_m^2+z_{pm}^2}{2z_{pm}r_p} \bigg)}
	\end{equation}
	
	\begin{equation}
		l_{2s} = \sqrt{r_p^2+z_{sp}^2-2r_pz_{sp}\cos \bigg(\cos^{-1} \frac{z_{sp}^2+z_{pm}^2-z_{sm}^2}{2z_{sp}z_{pm}} - \cos^{-1} \frac{r_p^2-r_m^2+z_{pm}^2}{2z_{pm}r_p} \bigg)}
	\end{equation}
	
	Similarly, the separation of the planet from the points of intersection of the moon and the star are given by,
	
	\begin{equation}
		l_{1p} = \sqrt{r_m^2+z_{pm}^2-2r_mz_{pm}\cos \bigg(\cos^{-1} \frac{z_{pm}^2+z_{sm}^2-z_{sp}^2}{2z_{pm}z_{sm}} + \cos^{-1} \frac{r_m^2-1+z_{sm}^2}{2z_{sm}r_m} \bigg)}
	\end{equation}
	
	\begin{equation}
		l_{2p} = \sqrt{r_m^2+z_{pm}^2-2r_mz_{pm}\cos \bigg(\cos^{-1} \frac{z_{pm}^2+z_{sm}^2-z_{sp}^2}{2z_{pm}z_{sm}} - \cos^{-1} \frac{r_m^2-1+z_{sm}^2}{2z_{sm}r_m} \bigg)}
	\end{equation}
	
	Also, the separation of the moon from the points of intersection of the star and the planet are given by,
	
	\begin{equation}
		l_{1m} = \sqrt{r_p^2+z_{pm}^2-2r_pz_{pm}\cos \bigg(\cos^{-1} \frac{z_{pm}^2+z_{sp}^2-z_{sm}^2}{2z_{pm}z_{sp}} + \cos^{-1} \frac{r_p^2-1+z_{sp}^2}{2z_{sp}r_p} \bigg)}
	\end{equation}
	
	\begin{equation}
		l_{2m} = \sqrt{r_p^2+z_{pm}^2-2r_pz_{pm}\cos \bigg(\cos^{-1} \frac{z_{pm}^2+z_{sp}^2-z_{sm}^2}{2z_{pm}z_{sp}} - \cos^{-1} \frac{r_p^2-1+z_{sp}^2}{2z_{sp}r_p} \bigg)}
	\end{equation}
	
	\begin{figure}
		\centering
		\includegraphics[width=0.6\linewidth]{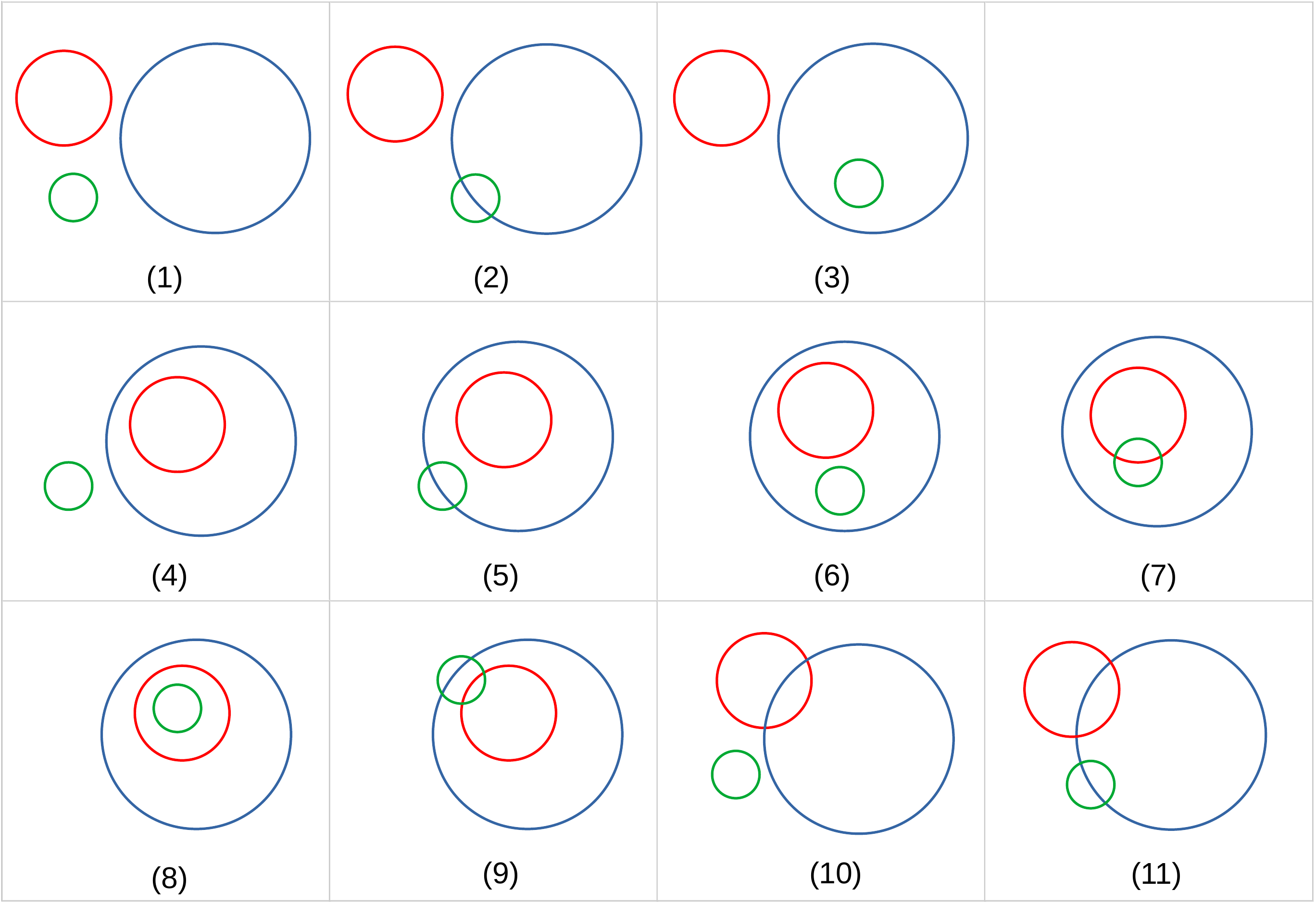}
		\includegraphics[width=0.6\linewidth]{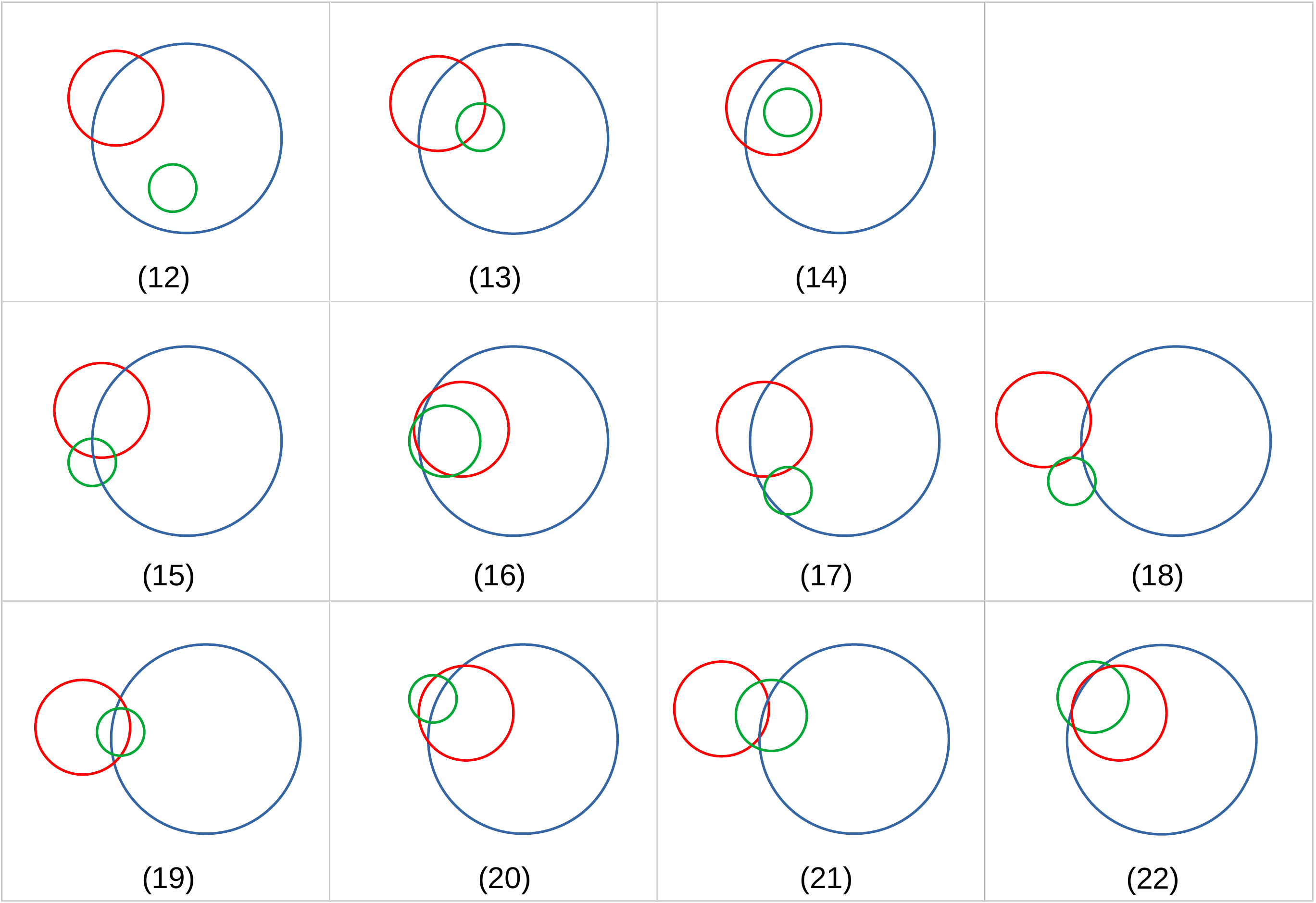}
		\caption{An instance of all possible cases of alignments for the star (blue), the planet (red) and the moon (green).}
		\label{fig:fig3}
	\end{figure}
	
	\begin{figure}
		\centering
		\includegraphics[width=0.5\linewidth]{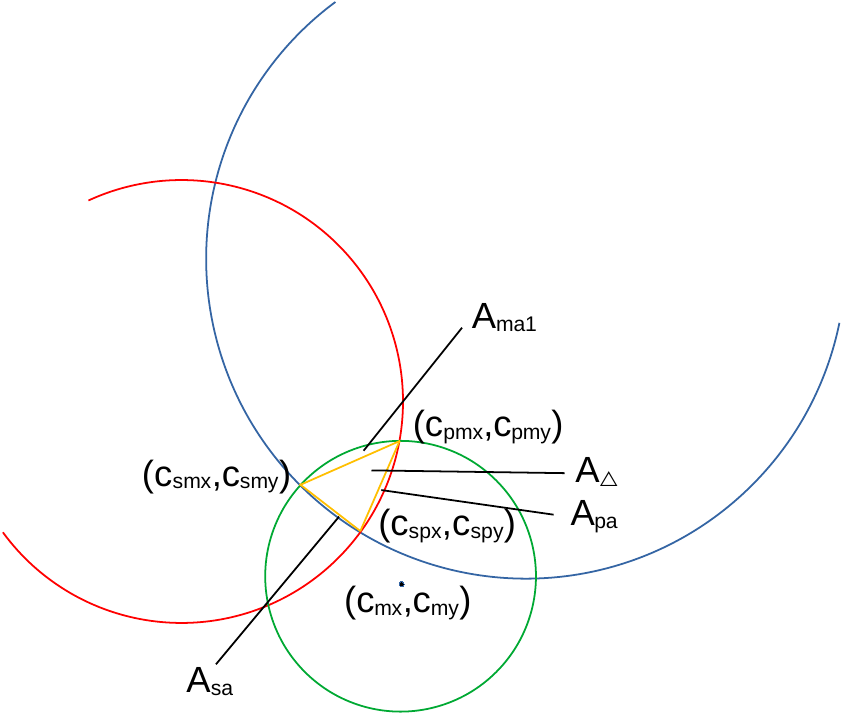}
		\caption{Alignment with intersection of all the three bodies (i.e., the star, the planet and the moon) showing $A_\triangle$, the area of the triangle and $A_{sa}$, $A_{pa}$ and $A_{ma1}$, the areas of the arcs within the common area of intersection of the three bodies.}
		\label{fig:fig2}
	\end{figure}
	
	The normalized flux from the star is given by,
	
	\begin{equation}
		F = 1 - \frac{F'}{F_T}
	\end{equation}

	where $F_T$ is the un-obscured flux of the star, and $F'$ is the occulted flux. If $F_p'$ is the flux occulted by the planet and $F_m'$ is that by the moon, then
	
	\begin{equation}
		F' = F_p' + F_m'
	\end{equation}
	
	Under small-planet approximation, i.e., $r_p, r_m\lesssim 0.1$, we follow the prescription by \citet{2002ApJ...580L.171M}. Hence, we have,
	
	\begin{equation}
		F_p' = \frac{2A_{op}}{(\rho_{2p}^2 - \rho_{1p}^2)} \int_{\rho_{1p}}^{\rho_{2p}} I(\rho) \rho d\rho
	\end{equation}
	
	\begin{equation}
		F_m' = \frac{2A_{om}}{(\rho_{2m}^2 - \rho_{1m}^2)} \int_{\rho_{1m}}^{\rho_{2m}} I(\rho) \rho d\rho
	\end{equation}
		
	\begin{equation}
		(\rho_{1p}, \rho_{2p}) = 
		\begin{cases}
			(0, z_{sp}+r_p), & z_{sp}\le r_p,\\
			(z_{sp}-r_p,z_{sp}+r_p), & r_p<z_{sp}<1-r_p,\\
			(z_{sp}-r_p,1), & 1-r_p \le z_{sp}<1+r_p
		\end{cases}
	\end{equation}
	
	\begin{equation}
		(\rho_{1m}, \rho_{2m}) = 
		\begin{cases}
			(0, z_{sm}+r_m), & z_{sm}\le r_m,\\
			(z_{sm}-r_m,z_{sm}+r_m), & r_m<z_{sm}<1-r_m,\\
			(z_{sm}-r_m,1), & 1-r_m \le z_{sm}<1+r_m
		\end{cases}
	\end{equation}

	where $I(\rho)$ is the specific intensity, $\rho$ being the radial distance; $A_{op}$ is the area of the stellar disk occulted by the planet; and $A_{om}$ is that occulted by the moon.
	
	Thus, the normalized flux from the star can be written as,
	
	\begin{equation}
		F = 1 - \frac{2}{F_T} \left[ \frac{A_{op}}{(\rho_{2p}^2 - \rho_{1p}^2)} \int_{\rho_{1p}}^{\rho_{2p}} I(\rho) \rho d\rho + \frac{A_{om}}{(\rho_{2m}^2 - \rho_{1m}^2)} \int_{\rho_{1m}}^{\rho_{2m}} I(\rho) \rho d\rho \right]
	\end{equation}

	Different alignments of the star, the planet and the moon result into different values of $A_{op}$ and $A_{om}$. We have categorized all the possible alignments into 22 cases, as shown in Figure \ref{fig:fig3}. Similar categorizations were previously provided by  \citet{fewell2006area} and \cite{2011MNRAS.416..689K}. However, we have made a more concise categorization considering only the physically feasible alignments and a more straightforward criteria for where they hold. The conditions where these cases hold, along with the values of $A_{op}$ and $A_{om}$ are listed in Table \ref{tab:tab2}. Various terms for the area used in Table \ref{tab:tab2} are as following.
	
	$A_p = \pi r_p^2$ and $A_m = \pi r_m^2$ are the disc areas of the planet and the moon respectively. $A_{sp}$, $A_{pm}$ and $A_{sm}$ are the areas of intersections of star-planet, planet-moon and star-moon respectively \citep{2002ApJ...580L.171M}, given by,
	
	\begin{equation}
		A_{sp} = cos^{-1} \frac{1-r_p^2+z_{sp}^2}{2z_{sp}} + r_p^2 cos^{-1} \frac{r_p^2-1+z_{sp}^2}{2z_{sp}r_p} - \sqrt{z_{sp}^2-\frac{(1-r_p^2+z_{sp}^2)^2}{4}}
	\end{equation}
	
	\begin{equation}
		A_{pm} = r_p^2 cos^{-1} \frac{r_p^2-r_m^2+z_{pm}^2}{2z_{pm}r_p} + r_m^2 cos^{-1} \frac{r_m^2-r_p^2+z_{pm}^2}{2z_{pm}r_m} - \sqrt{z_{pm}^2r_p^2- \frac{(r_p^2-r_m^2+z_{pm}^2)^2}{4}}
	\end{equation}
	
	\begin{equation}
		A_{sm} = cos^{-1} \frac{1-r_m^2+z_{sm}^2}{2z_{sm}} + r_m^2 cos^{-1} \frac{r_m^2-1+z_{sm}^2}{2z_{sm}r_m} - \sqrt{z_{sm}^2-\frac{(1-r_m^2+z_{sm}^2)^2}{4}}
	\end{equation}
	
	Following a similar formalism as given by \citet{fewell2006area}, the area of intersection of all the three bodies, i.e., the star, the planet and the moon, (see Figure \ref{fig:fig2}) can be written as,
	
	\begin{equation}
		A_{spm1} = A_\triangle + A_{sa} + A_{pa} + A_{ma1}
	\end{equation}

	when $D\ge 0$. Otherwise,
		
	\begin{equation}
		A_{spm2} = A_\triangle + A_{sa} + A_{pa} + A_{ma2}
	\end{equation}

	where, 
	
	\begin{equation}
		D = (c_{mx}-c_{pmx}) (c_{smy}-c_{pmy}) - (c_{my}-c_{pmy}) (c_{smx}-c_{pmx}).
	\end{equation}

	which determines whether more than half of the moon is within the arc area (see Figure \ref{fig:fig2}).

	In the above expressions,
	
	\begin{equation}
		A_\triangle = \sqrt{s(s-s_{sp})(s-s_{pm})(s-s_{sm})},
	\end{equation}
	\begin{equation}
		A_{sa} = sin^{-1} \frac{s_{pm}}{2} - \frac{s_{pm}}{4} \sqrt{4-s_{pm}^2},
	\end{equation}
	\begin{equation}
		A_{pa} = r_p^2 sin^{-1} \frac{s_{sm}}{2r_p} - \frac{s_{sm}}{4} \sqrt{4r_p^2-s_{sm}^2},
	\end{equation}
	\begin{equation}
		A_{ma1} = r_m^2 sin^{-1} \frac{s_{sp}}{2r_m} - \frac{s_{sp}}{4} \sqrt{4r_m^2-s_{sp}^2},
	\end{equation}
	\begin{equation}
		A_{ma2} = r_m^2 sin^{-1} \frac{s_{sp}}{2r_m} + \frac{s_{sp}}{4} \sqrt{4r_m^2-s_{sp}^2},
	\end{equation}
	\begin{equation}
		s_{sp} = \sqrt{(c_{smx}-c_{pmx})^2 + (c_{smy}-c_{pmy})^2},
	\end{equation}
	\begin{equation}
		s_{pm} = \sqrt{(c_{spx}-c_{smx})^2 + (c_{spy}-c_{smy})^2},
	\end{equation}
	\begin{equation}
		s_{sm} = \sqrt{(c_{spx}-c_{pmx})^2 + (c_{spy}-c_{pmy})^2},
	\end{equation}	
	\begin{equation}
		s = \frac{s_{sp}+s_{pm}+s_{sm}}{2},
	\end{equation}	
	\begin{equation}
		(c_{mx},c_{my}) = (z_{sm}\cos \delta_{sm}, -z_{sm} \sin \delta_{sm}),
	\end{equation}
	\begin{equation}
		\cos \delta_{sm} = \frac{z_{sp}^2+z_{sm}^2-z_{pm}^2}{2z_{sp}z_{sm}},
	\end{equation}
	\begin{equation}
		(c_{spx},c_{spy}) = \Bigg[\frac{1-r_p^2+z_{sp}^2}{2z_{sp}}, -\sqrt{1-\bigg(\frac{1-r_p^2+z_{sp}^2}{2z_{sp}}\bigg)^2}\Bigg],
	\end{equation}
	\begin{equation}
		(c_{smx},c_{smy}) = (c_{smx}'\cos \delta_{sm} + c_{smy}'\sin \delta_{sm}, -c_{smx}'\sin \delta_{sm} + c_{smy}'\cos \delta_{sm}),
	\end{equation}
	\begin{equation}
		(c_{smx}',c_{smy}') = \Bigg[\frac{1-r_m^2+z_{sm}^2}{2z_{sm}}, \sqrt{1-\bigg(\frac{1-r_m^2+z_{sm}^2}{2z_{sm}}\bigg)^2}\Bigg],
	\end{equation}
	\begin{equation}
		(c_{pmx},c_{pmy}) = (-c''_{pmx}\cos \delta_{pm} + c''_{pmy}\sin \delta_{pm} -z_{sp}, -c''_{pmx}\sin \delta_{pm} + c''_{pmy}\cos \delta_{pm}),
	\end{equation}
	\begin{equation}
		(c''_{pmx},c''_{pmy}) = \Bigg[\frac{r_p-r_m^2+z_{pm}^2}{2z_{pm}}, -\sqrt{r_p^2-\bigg(\frac{r_p^2-r_m^2+z_{pm}^2}{2z_{pm}}\bigg)^2}\Bigg],
	\end{equation}
	and
	\begin{equation}
		\cos \delta_{pm} = \frac{z_{sp}^2+z_{pm}^2-z_{sm}^2}{2z_{sp}z_{pm}}.
	\end{equation}
	
	Here, $A_\triangle$ is the area of the triangle and $A_{sa}$, $A_{pa}$, $A_{ma1}$ and $A_{ma2}$ are the areas of the arcs within the area of intersection of the three bodies (i.e., the star, the planet and the moon); $c_{sp}$, $c_{pm}$ and $c_{sm}$ are the coordinates of intersection of the three bodies; and $c_m$ is the coordinate of the center of the moon.
	 	
	\startlongtable
	\begin{deluxetable}{CCCC}
		\tablecaption{Different cases of star-planet-moon alignments with the conditions at where they hold and the values for $A_{op}$ and $A_{om}$, the areas of the stellar disk occulted by the planet and   the moon respectively.}
		\label{tab:tab2}
		\tablehead{\colhead{Case} & \colhead{Conditions} & \colhead{$\mathrm{A_{op}}$} & \colhead{$\mathrm{A_{om}}$}}
		\startdata
		1 & z_{sp}\ge 1+r_p & 0 & 0 \\
		& z_{sm}\ge 1+r_m \\
		\hline
		2 & z_{sp}\ge 1+r_p & 0 & A_{sm}\\
		& 1+r_m>z_{sm}>1-r_m \\
		\hline
		3 & z_{sp}\ge 1+r_p & 0 & A_{pm}\\
		& z_{sm}\le 1-r_m \\
		\hline
		4 & z_{sp}\le 1-r_p & A_p & 0\\
		& z_{pm}\ge r_p+r_m\\
		& z_{sm}\ge 1+r_m\\
		\hline
		5 & z_{sp}\le 1-r_p & A_p & A_{sm}\\
		& z_{pm}\ge r_p+r_m \\
		& 1+r_m>z_{sm}>1-r_m \\
		\hline
		6 & z_{sp}\le 1-r_p & A_p & A_m\\
		& z_{pm}\ge r_p+r_m \\
		& z_{sm}\le 1-r_m \\
		\hline
		7 & z_{sp}\le 1-r_p & A_p & A_m-A_{pm}\\
		& r_p+r_m>z_{pm}>r_p-r_m\\
		& z_{sm}\le 1-r_m \\
		\hline
		8 & z_{sp}\le 1-r_p & A_p & 0\\
		& z_{pm}\le r_p-r_m\\
		& z_{sm}\le 1-r_m \\
		\hline
		9 & z_p\le 1-r_{sp} & A_p & A_{sm}-A_{pm}\\
		& r_p+r_m>z_{pm}>r_p-r_m\\
		& 1+r_m>z_{sm}>1-r_m \\
		\hline
		10 & 1+r_p>z_{sp}>1-r_p & A_{sp} & 0\\
		& z_{sm}\ge 1+r_m\\
		\hline
		11 & 1+r_p>z_{sp}>1-r_{sp} & A_{sp} & A_{sm}\\
		& z_m\ge r_p+r_m\\
		& 1+r_m>z_{sm}>1-r_m \\
		\hline
		12 & 1+r_p>z_{sp}>1-r_p & A_{sp} & A_m\\
		& z_{pm}\ge r_p+r_m\\
		& z_{sm}\le 1-r_m \\
		\hline
		13 & 1+r_p>z_{sp}>1-r_p & A_{sp} & A_m-A_{pm}\\
		& r_p+r_m>z_{pm}>r_p-r_m\\
		& z_{sm}\le 1-r_m \\
		\hline
		14 & 1+r_p>z_{sp}>1-r_p & A_{sp} & 0\\
		& z_{pm}\le r_p-r_m\\
		\hline
		15 & 1+r_p>z_{pm}>1-r_p & A_{sp} & A_{sm}-A_{spm1}\\
		& r_p+r_m>z_{pm}>r_p-r_m \\
		& 1+r_m>z_{sm}>1-r_m \\
		& l_{1m}>r_m>l_{2m} \\
		& D\ge 0 \\
		\hline
		16 & 1+r_p>z_{sp}>1-r_p & A_{sp} & A_{sm}-A_{spm2}\\
		& r_p+r_m>z_{pm}>r_p-r_m \\
		& 1+r_m>z_{sm}>1-r_m \\
		& l_{1m}>r_m>l_{2m} \\
		& D<0 \\
		\hline
		17 & 1+r_p>z_{sp}>1-r_p & A_{sp} & A_{sm}-A_{pm}\\
		& r_p+r_m>z_{pm}>r_p-r_m \\
		& 1+r_m>z_{sm}>1-r_m \\
		& 1\ge l_{1s}\ge l_{2s} \\
		& l_{1p}\ge l_{2p}\ge r_p \\
		& l_{1m}\ge l_{2m}\ge r_m \\
		\hline
		18 & 1+r_p>z_{pm}>1-r_p & A_{sp} & A_{sm}\\
		& r_p+r_m>z_{pm}>r_p-r_m \\
		& 1+r_m>z_{sm}>1-r_m \\
		& l_{1s}\ge l_{2s}\ge 1 \\
		& l_{1p}\ge l_{2p}\ge r_p \\
		& l_{1m}\ge l_{2m}\ge r_m \\
		\hline
		19 & 1+r_p>z_{sp}>1-r_p & A_{sp} & A_m-A_{pm}\\
		& r_p+r_m>z_{pm}>r_p-r_m \\
		& 1+r_m>z_{sm}>1-r_m \\
		& 1\ge l_{1s}\ge l_{2s} \\
		& r_p\ge l_{1p}\ge l_{2p} \\
		& l_{1m}\ge l_{2m}\ge r_m \\
		\hline
		20 & 1+r_p>z_{sp}>1-r_p & A_{sp} & 0\\
		& r_p+r_m>z_{pm}>r_p-r_m \\
		& 1+r_m>z_{sm}>1-r_m \\
		& l_{1s}\ge l_{2s}\ge 1 \\
		& r_p\ge l_{1p}\ge l_{2p} \\
		& l_{1m}\ge l_{2m}\ge r_m \\
		\hline
		21 & 1+r_p>z_{sp}>1-r_p & A_{sp} & A_{sm}-A_{sp}\\
		& r_p+r_m>z_{pm}>r_p-r_m \\
		& 1+r_m>z_{sm}>1-r_m \\
		& l_{1s}\ge l_{2s}\ge 1 \\
		& l_{1p}\ge l_{2p}\ge r_p \\
		& r_m\ge l_{1m}\ge l_{2m} \\
		\hline
		22 & 1+r_p>z_{sp}>1-r_p & A_{sp} & A_{sm}-A_{pm}+(A_p-A_{sp})\\
		& r_p+r_m>z_{sp}>r_p-r_m \\
		& 1+r_m>z_{sm}>1-r_m \\
		& 1\ge l_{1s}\ge l_{2s} \\
		& l_{1p}\ge l_{2p}\ge r_p \\
		& r_m\ge l_{1m}\ge l_{2m} \\
		\enddata
	\end{deluxetable}
	
	\section{Results and discussion}\label{sec3}
	
	The model parameters in the present formalism are $r_p$, $r_m$, $t_{0b}$, $t_{0m}$, $P_b$, $P_m$, $r_{pm}$, $M$, $i_b$, $i_m$, $\alpha_{mb}$, and the limb-darkening coefficients of the host-star defining $I(\rho)$, where all the distances are in terms of stellar radius. Here we have used the quadratic limb-darkening formula \citep{claret1990limb, claret2000new}, given by
	
	\begin{equation}
		I(\rho)/I(0) = 1 - u_1 (1-\mu) - u_2 (1-\mu)^2,
	\end{equation}

	where $\mu = \sqrt{1-\rho ^2}$, and $u_1$ and $u_2$ are the quadratic limb-darkening coefficients.
	
	While generating the model transit lightcurves, we have used the analytical quadratic limb-darkening formalism for transit flux given by \citet{2002ApJ...580L.171M}. In the absence of such an analytical quadratic limb-darkening formula, we have used the small-body approximation to estimate the transit flux for the cases where all the three circular bodies overlap  (see Section \ref{sec2}).
	
	\begin{figure}
		\centering
		\includegraphics[width=1.0\linewidth]{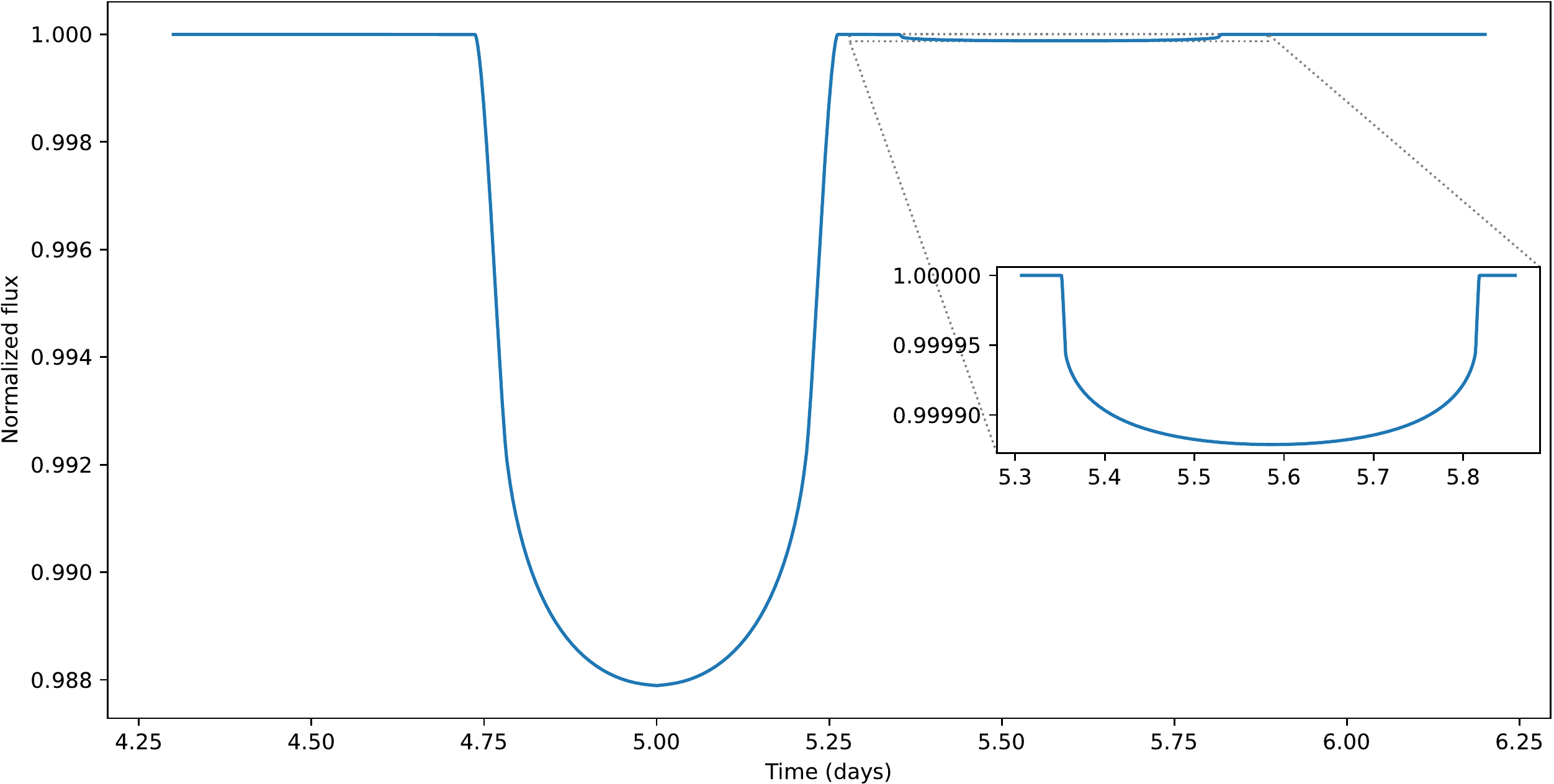}
		\text{(a)}
		\includegraphics[width=1.0\linewidth]{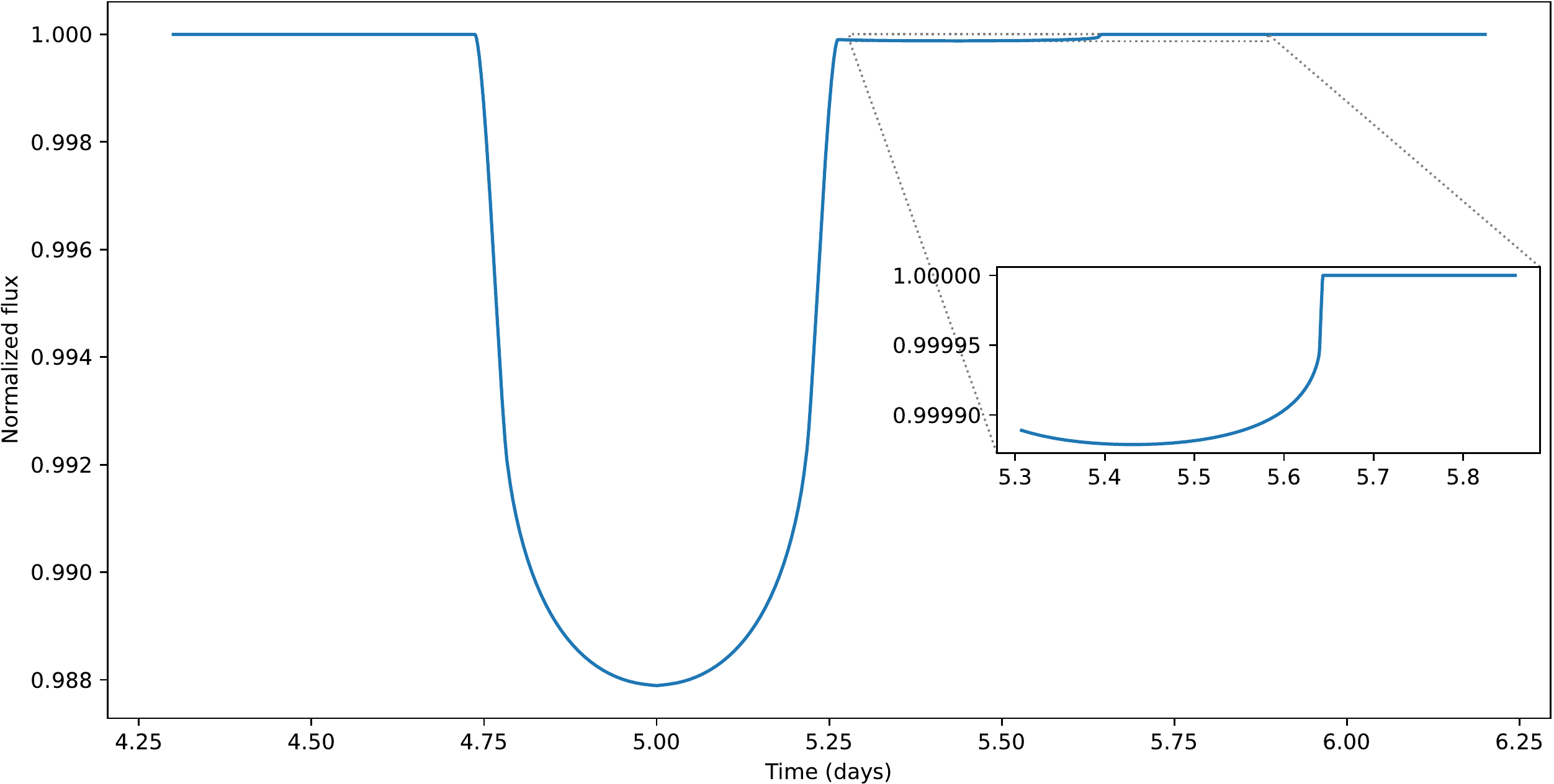}
		\text{(b)}
		\caption{Transit light-curves of a moon hosting exoplanetary system: (a) with $r_p = 0.1$, $r_m = 0.01$, $t_{0b} = 5$ days, $t_{0m} = 10$ days, $P_b = 300$ days, $P_m = 20$ days, $r_{pm} = 200$, $M_p/M_m = 1411$, $i_b = 90^o$, $i_m = 90^o$, $\alpha_{mb} = 0^o$, $u_1 = 0.4$ and $u_2 = 0.25$; and (b) replacing $t_{0m} = 8$ days.}
		\label{fig:fig4}
	\end{figure}
	
	\begin{figure}
		\centering
		\includegraphics[width=1.0\linewidth]{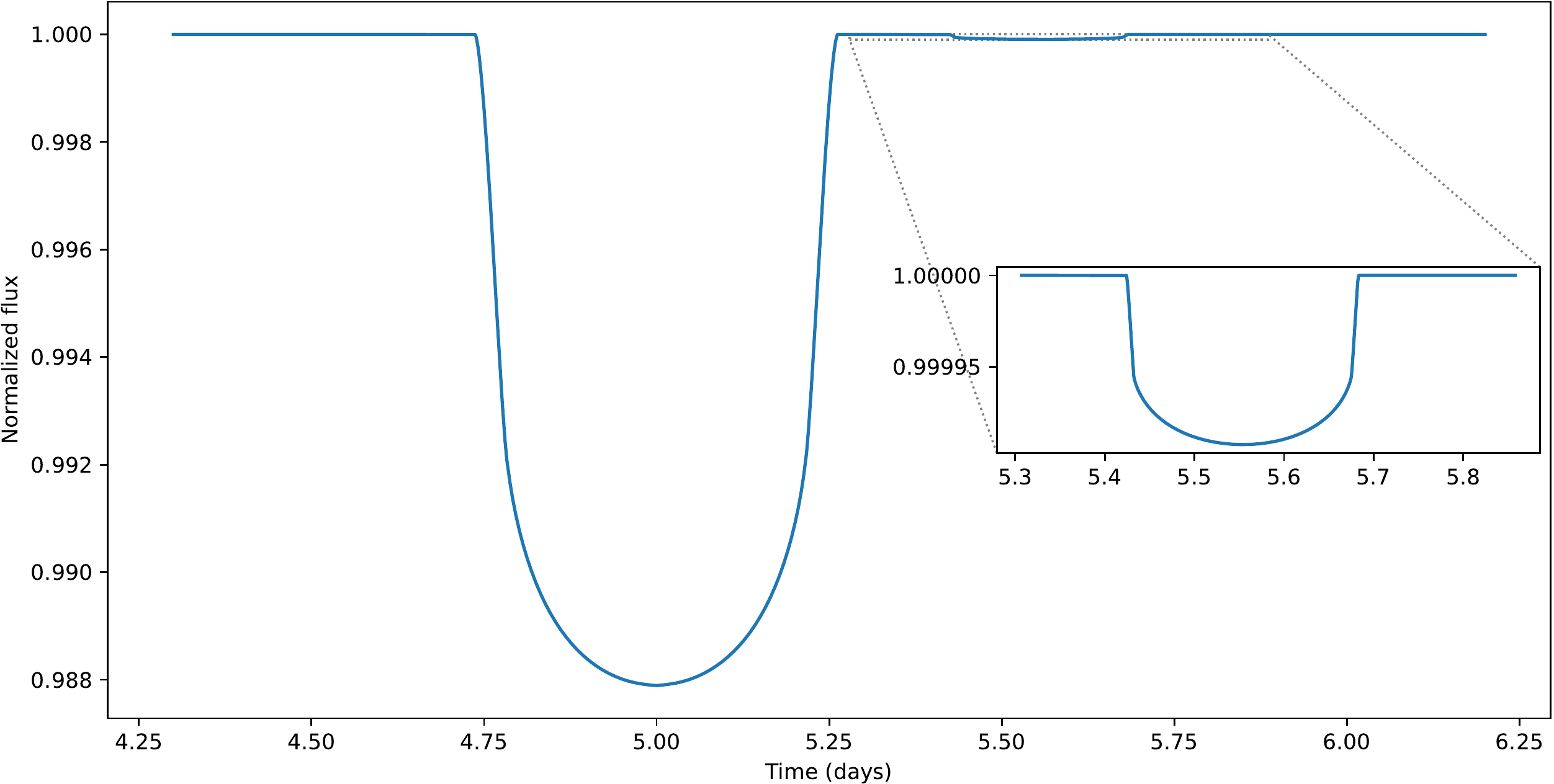}
		\text{(a)}
		\includegraphics[width=1.0\linewidth]{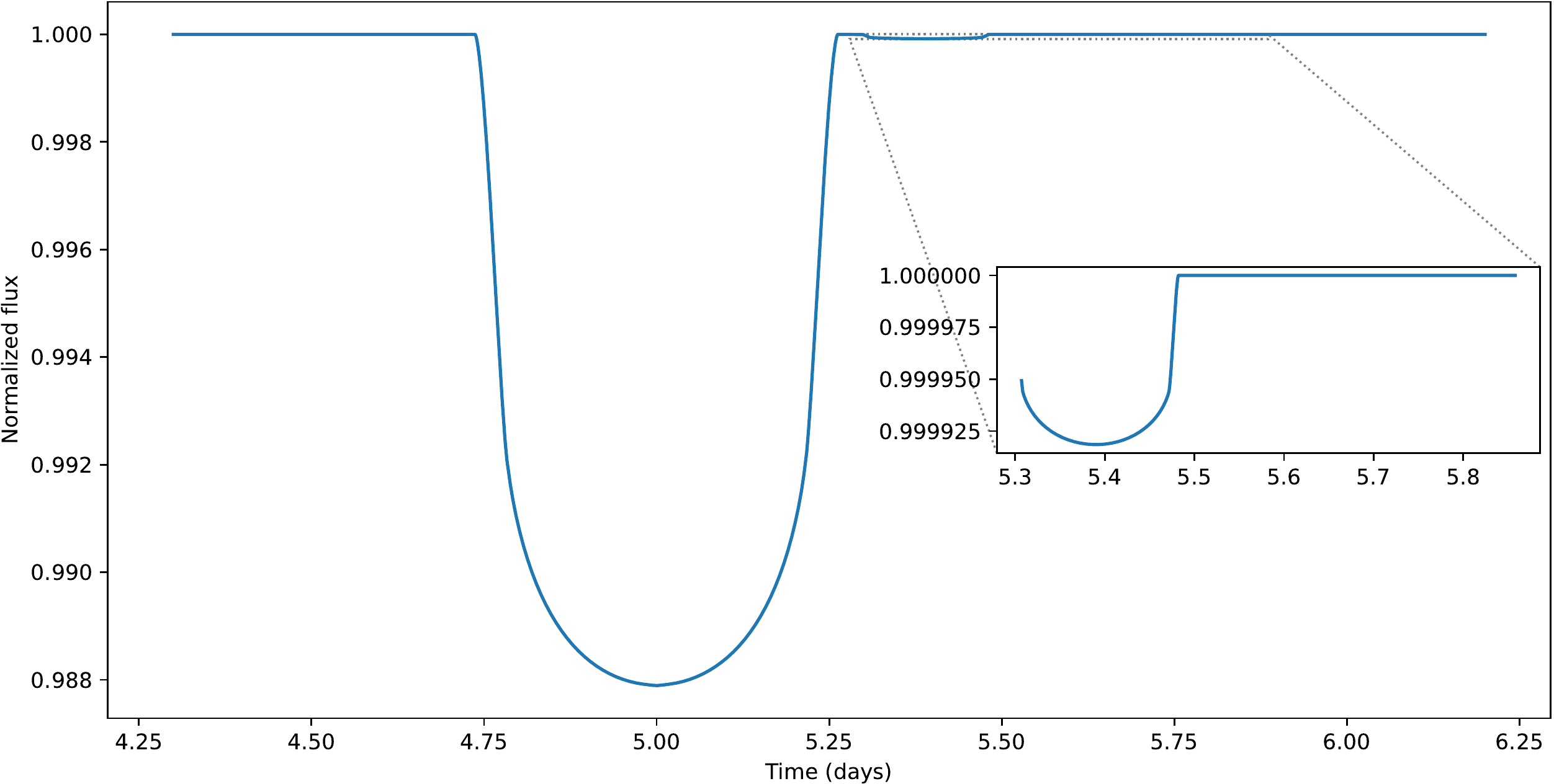}
		\text{(b)}
		\caption{Transit light-curves of a moon hosting exoplanetary system: (a) with $r_p = 0.1$, $r_m = 0.01$, $t_{0b} = 5$ days, $t_{0m} = 10$ days, $P_b = 300$ days, $P_m = 20$ days, $r_{pm} = 200$, $M_p/M_m = 1411$, $i_b = 90^o$, $i_m = 90^o$, $\alpha_{mb} = 20^o$, $u_1 = 0.4$ and $u_2 = 0.25$; and (b) replacing $t_{0m} = 8$ days and $\alpha_{mb} = 30^o$.}
		\label{fig:fig5}
	\end{figure}
	
	\begin{figure}
		\centering
		\includegraphics[width=1.0\linewidth]{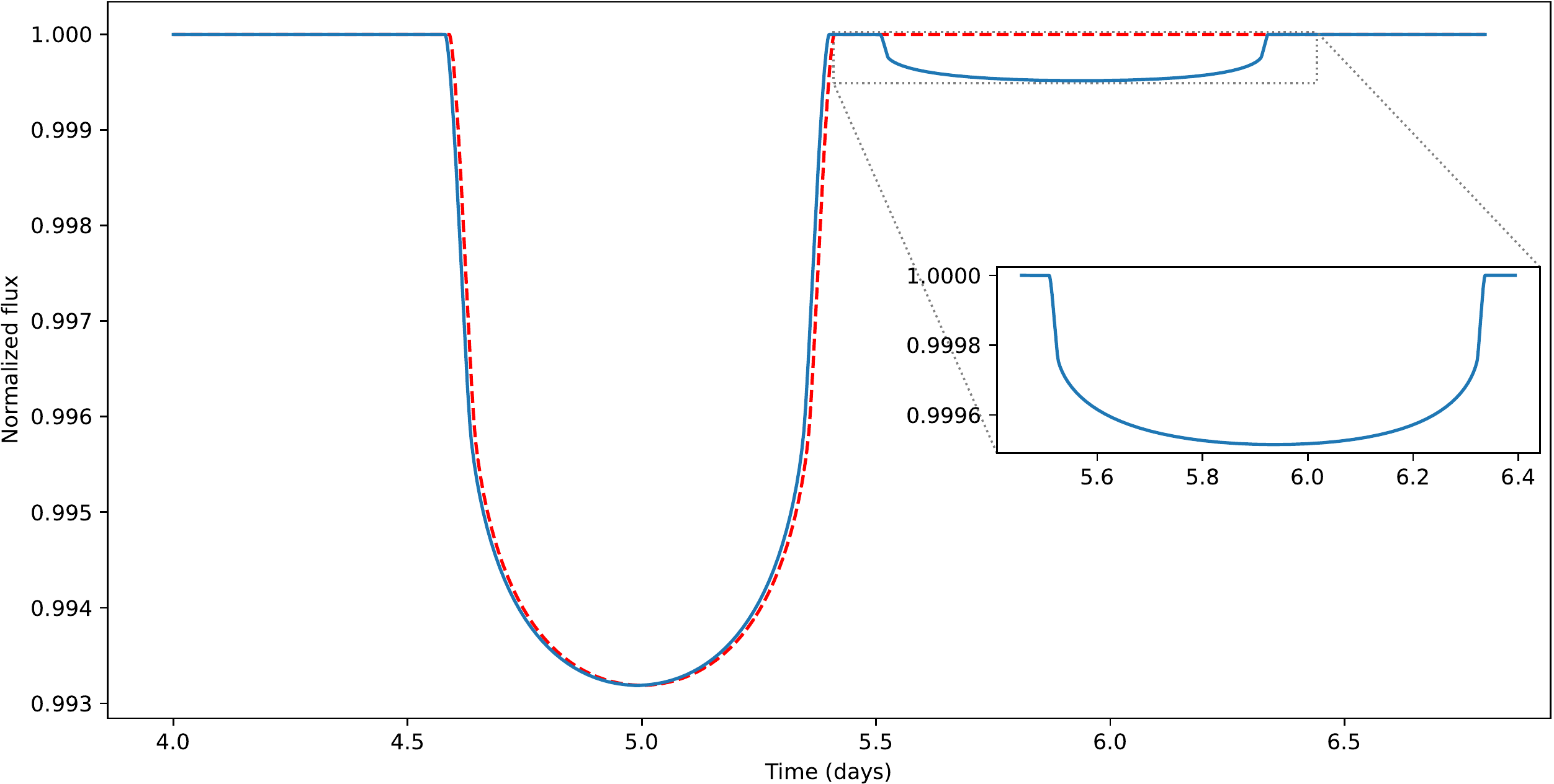}
		\text{(a)}
		\includegraphics[width=1.0\linewidth]{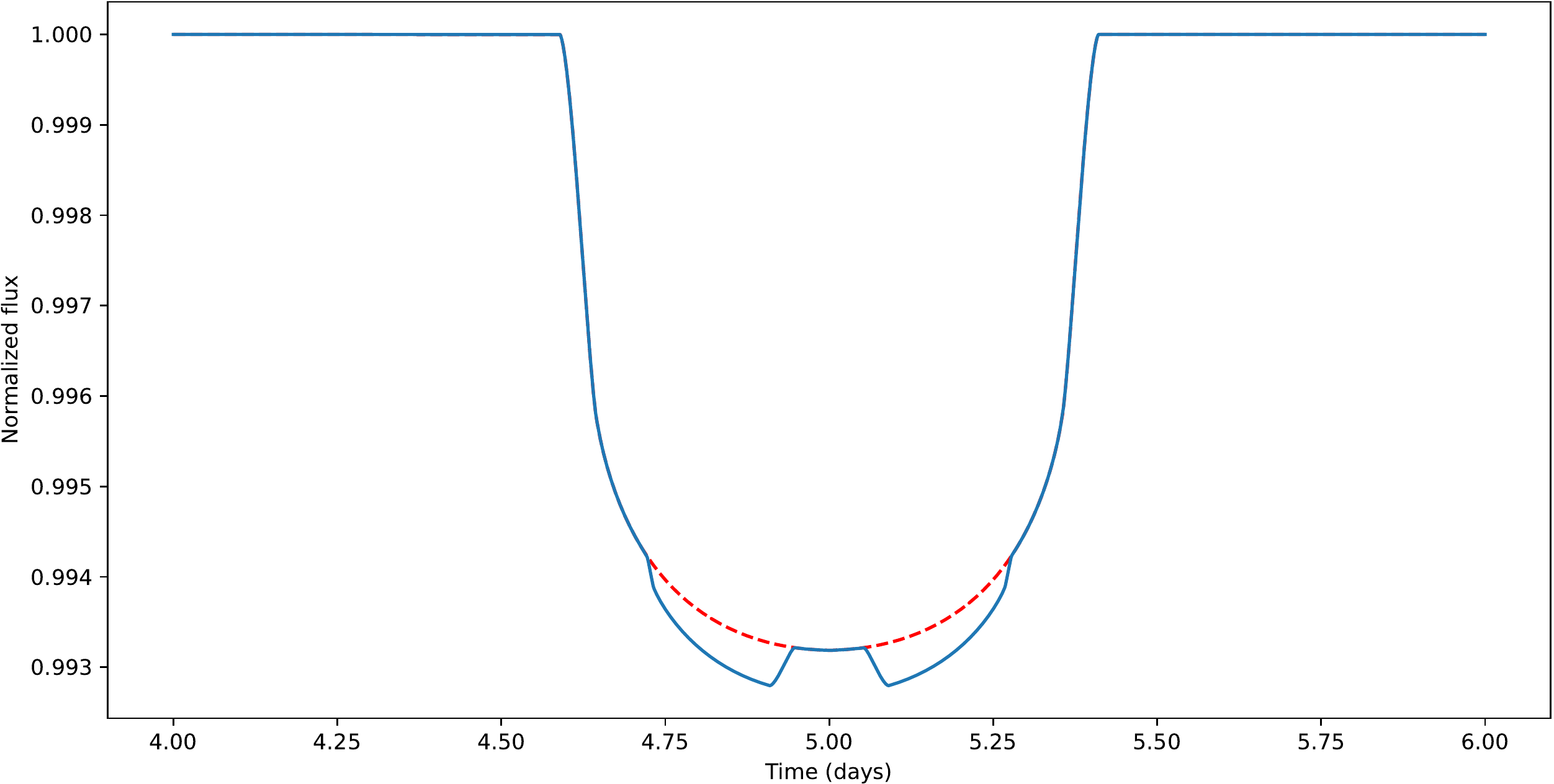}
		\text{(b)}
		\caption{Transit light-curves of a moon hosting exoplanetary system: (a) with $r_p = 0.075$, $r_m = 0.02$, $t_{0b} = 5$ days, $t_{0m} = 10$ days, $P_b = 60$ days, $P_m = 15$ days, $r_{pm} = 25$, $M_p/M_m = 81$, $i_b = 90^o$, $i_m = 90^o$, $\alpha_{mb} = 0^o$, $u_1 = 0.4$ and $u_2 = 0.25$; and (b) same but with $t_{0m} = 5$ days. The dashed red lines show the transit lightcurves of the planet in the absence of a moon. The transit-time-variation (TTV) due to the presence of a moon can be observed in (a).}
		\label{fig:fig6}
	\end{figure}
	
	Let us now consider a scenario with $r_p = 0.1$, $r_m = 0.01$, $t_{0b} = 5$ days, $t_{0m} = 10$ days, $P_b = 300$ days, $P_m = 20$ days, $r_{pm} = 200$, $M_p/M_m = 1411$, $i_b = 90^o$, $i_m = 90^o$, $\alpha_{mb} = 0^o$, $u_1 = 0.4$ and $u_2 = 0.25$. Clearly, $i_b = i_m= 90^o$  implies that both the planet and the moon are transiting through the center of the star, and that combined with $\alpha_{mb} = 0^o$ implies that the orbit of the moon is aligned with the orbit of the planet, i.e., both the planet and the moon are in the same orbital plane. The transit lightcurve for this scenario is shown in Figure \ref{fig:fig4}(a). Usually, the transit of the moon can take place before, during or after the transit of the planet depending upon a combination of various parameters. For this scenario, we can see that the transit of the moon starts after the end of the planet's transit, as the moon is placed in a wide orbit around the planet and its position makes it highly trailing while transiting the star. If we change the position of the moon by replacing $t_{0m} = 8$ days, as shown in Figure \ref{fig:fig4}(b), we can see that the transit of the moon starts before the end of the transit of the planet.

	The alignment of the orbit of an exomoon depends upon the formation and evolution path it followed. If the moon is formed from the circumplanetary disc, there is a higher probability for its orbit to be equatorial and it might be co-aligned (co-planer) with the planetary orbit \citep{1999ARA&A..37..533P}. On the other hand, if the moon is formed through planetary capture or collision, its orbit may not be co-aligned with the orbit of the planet. Both these situations can easily be modeled by using our formalism. When $i_b = i_m$ and $\alpha_{mb} = 0$, the orbits are co-aligned. On the contrary, when $i_b \ne i_m$ and/or $\alpha_{mb} \ne 0$, the orbits are no longer co-aligned with each other. To demonstrate it, lets consider a scenario by replacing $\alpha_{mb} = 20^o$ in the first scenario. As shown in Figure \ref{fig:fig5}(a), we can see that both the transit depth and duration of the moon has decreased. This is because, in this case the moon is transiting towards the edge of the star instead of through the center as was in the case of the first scenario.
	
	Obviously, if the moon is in a position such that it is fully transiting the planet or is fully eclipsed by it, while transiting the star, no transit signal due to the moon could be observed. Also, if the moon is in a wide and highly inclined orbit as compared to that of the planet, it may not transit the star every time the planet transits it \citep{2019ApJ...875L..25M}. For example, if we replace $\alpha_{mb} = 30^o$ in the previous case, no transit is observed for the moon. However, if the position of the moon is changed by replacing $t_{0m} = 8$ days, the transit of the moon is observed as shown in Figure \ref{fig:fig5}(b). Combining these factors along with the fact that exomoons are more likely to be found around planets in wider orbits around their host-stars, i.e. planets with much longer orbital period than a few days, it would require long period surveys continuously monitoring a particular portion of the sky to detect the exomoons. However, such surveys are also likely to increase the number of large period exoplanets, including the habitable-zone terrestrial exoplanets, thereby increasing their effectiveness by detecting many interesting planetary and sub-planetary mass bodies.
	
	The photometric precision required to detect the exomoons is directly related to the transit depth, which is in turn dependent upon the ratio of disc area of the moon to that of the star. The required photometric precision is minimum for an exomoon in a system with a smaller M-dwarf type star. Lets consider such a system with a moon of the size of the Moon around a planet of the size of the Earth, i.e. $r_p = 0.075$, $r_m = 0.02$, $t_{0b} = 5$ days, $t_{0m} = 10$ days, $P_b = 60$ days, $P_m = 15$ days, $r_{pm} = 25$, $M_p/M_m = 81$, $i_b = 90^o$, $i_m = 90^o$, $\alpha_{mb} = 0^o$, $u_1 = 0.4$ and $u_2 = 0.25$, the light-curve for which is shown in Figure \ref{fig:fig6}(a). If we change the position of the moon by replacing $t_{0m} = 5$ days, there would arise a scenario where the moon transits the planet while simultaneously transiting the star as well. Such a scenario is presented in Figure \ref{fig:fig6}(b).
	
	To demonstrate the affect of the moon on the transit timing of the planet, we have co-plotted the transit lightcurve of the planet in the absence of a moon in Figure \ref{fig:fig6}(a). Comparing the two lightcurves, the transit-time-variation (TTV) can be observed. For a practical scenario, it could be difficult to detect the TTV in the observed transit data, as the barycentric offset from the center of the planet is quite small compared to the distance between the centers of the planet and the moon, even for a smaller mass ratio between the planet and the moon (e.g. the earth-moon system, where $M_p/M_m \simeq 81$).
	
	On the other hand,  a higher precision would be required for systems with a larger host-star. The minimum photometric precision required to detect a terrestrial exoplanet of the size of the Earth around a star similar to the Sun is about 100 ppm (parts-per-million). Therefore,  a precision much better than that would be required to detect the exomoons around such systems. Such extremely high precision is expected to be achievable using the next generation large telescopes, such as the James Webb Space Telescope (JWST), the European Extremely Large Telescope (E-ELT), the Thirty Meter Telescope (TMT), and the Giant Magellan Telescope (GMT) etc. Also, the instrumental and atmospheric noise effects has to be minimized for such observations.  This can be achieved by using small-scale noise reduction techniques such as the wavelet denoising \citep{Donoho1994IdealDI, 2019AJ....158...39C, 2021AJ....162...18S, 2021AJ....162..221S}. The stellar variability and pulsations can also cause a challenge in such observations, which need to be reduced using techniques like the Gaussian process regression \citep{2006gpml.book.....R, 2015ApJ...810L..23J, 2019AJ....158...39C, 2021AJ....162...18S, 2021AJ....162..221S}.
	
	\section{Conclusion}\label{sec4}
	
	In this paper, we have presented an analytical formalism to model the transit lightcurves for a system with a transiting exoplanet hosting an exomoon. The formalism uses the radius and orbital properties of both the planet and the moon as model parameters. The orbital alignment of the  moon is taken care by introducing two angular parameters and hence both co-aligned and non-coaligned orbit of a moon with respect to the planetary orbit can be modeled easily. This also enables to model every possible scenarios of alignments for the star-planet-moon system using this formalism.
	
	The detection of exomoons requires extremely high precision observations which is expected to be achievable using the next generation very large telescopes along with the implementation of the existing critical noise reduction techniques. In such possibilities, our transit formalism could be useful to model the lightcurves in order to characterize the physical properties of the exomoons as well as to simulate every possible scenarios and make strategies for such extremely time-critical observations.
	
	We are thankful to the anonymous reviewer for a critical reading of the manuscript and for providing many useful comments and suggestions.
	
	\bibliography{ms}{}
	\bibliographystyle{aasjournal}
	
	\appendix
	
	\section{Derivation of \lowercase{$l_{1s}$} and \lowercase{$l_{2s}$}}\label{sec7}
	
	\begin{figure}[h]
		\centering
		\includegraphics[width=0.4\linewidth]{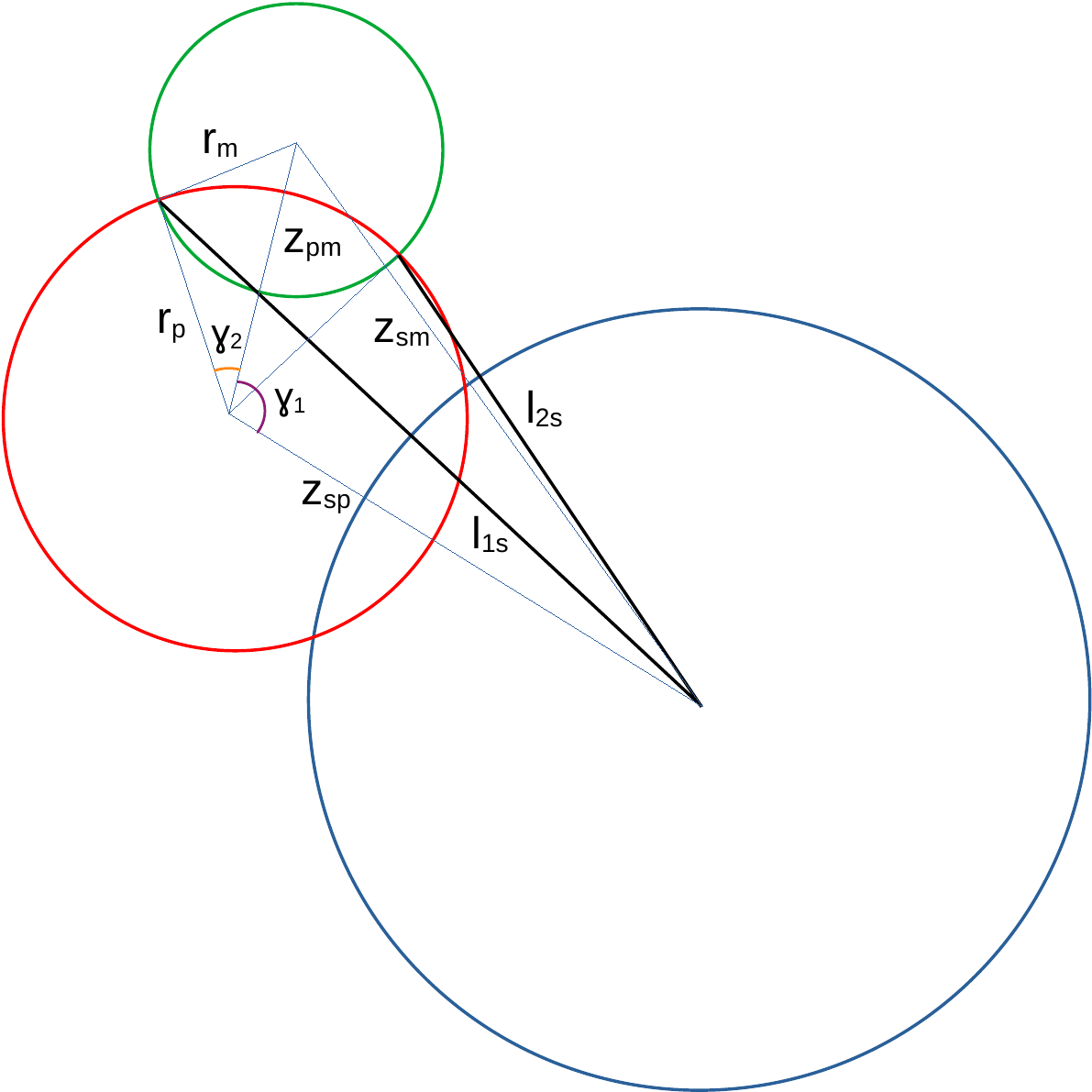}
		\caption{Alignment of the three bodies with star-planet and planet-moon intersections.}
		\label{fig:fig7}
	\end{figure}
	
	From Figure \ref{fig:fig7}, the angles $\gamma_1$ and $\gamma_2$ can be written as,
	
	\begin{equation}
		\gamma_1 = cos^{-1} \bigg(\frac{z_{sp}^2+z_{pm}^2-z_{sm}^2}{2z_{sp}z_{pm}}\bigg)
	\end{equation}
	
	\begin{equation}
		\gamma_2 = cos^{-1} \bigg(\frac{r_p^2-r_m^2+z_{pm}^2}{2z_{pm}r_p}\bigg)
	\end{equation}
	
	Now, $l_{1s}$ and $l_{2s}$ can be written as,
	
	\begin{align}
		l_{1s} &= \sqrt{r_p^2+z_{sp}^2-2r_pz_{sp}\cos(\gamma_1 + \gamma_2) } \nonumber\\
		&= \sqrt{r_p^2+z_{sp}^2-2r_pz_{sp}\cos \bigg(\cos^{-1} \frac{z_{sp}^2+z_{pm}^2-z_{sm}^2}{2z_{sp}z_{pm}} + \cos^{-1} \frac{r_p^2-r_m^2+z_{pm}^2}{2z_{pm}r_p} \bigg)}
	\end{align}
	
	\begin{align}
		l_{2s} &= \sqrt{r_p^2+z_{sp}^2-2r_pz_{sp}\cos(\gamma_1 - \gamma_2) } \nonumber\\
		&= \sqrt{r_p^2+z_{sp}^2-2r_pz_{sp}\cos \bigg(\cos^{-1} \frac{z_{sp}^2+z_{pm}^2-z_{sm}^2}{2z_{sp}z_{pm}} - \cos^{-1} \frac{r_p^2-r_m^2+z_{pm}^2}{2z_{pm}r_p} \bigg)}
	\end{align}
	
	$l_{1p}$, $l_{2p}$, $l_{1m}$ and $l_{2m}$ can also be derived in a similar fashion.
	
\end{document}